\documentstyle[aps,preprint,pra]{revtex}
\tighten
\begin{document}
\draft
\title{Making An Empty Promise With A Quantum Computer\footnote{This manuscript
 is written for a special issue on quantum computation in Fortschritte
 der Phys.}}
\author{H. F. Chau$^1$\footnote{e-mail: hfchau@hkusua.hku.hk} and Hoi-Kwong
 Lo$^2$\footnote{e-mail: hkl@hplb.hpl.hp.com}}
\address{$^1$ Department of Physics, University of Hong Kong, Pokfulam Road,
 Hong Kong}
\address{$^2$ Hewlett-Packard Labs, Filton Road, Stoke Gifford, Bristol
 BS12~6QZ, United Kingdom}
\date{\today}
\preprint{HKUPHYS-HFC-03; quant-ph/9709053}
\maketitle
\begin{abstract}
 Alice has made a decision in her mind. While she does not want to reveal it to
 Bob at this moment, she would like to convince Bob that she is committed to
 this particular decision and that she cannot change it at a later time. Is
 there a way for Alice to get Bob's trust? Until recently, researchers had
 believed that the above task can be performed with the help of quantum
 mechanics. And the security of the quantum scheme lies on the uncertainty
 principle. Nevertheless, such optimism was recently shattered by Mayers and by
 us, who found that Alice can always change her mind if she has a quantum
 computer. Here, we survey this dramatic development and its implications on
 the security of other quantum cryptographic schemes.
\end{abstract}
\medskip
\section{Introduction}
\label{S:Intro}
 Cryptography --- the art of sending secret messages --- has a long and
 distinguished history of applications. The security of conventional
 cryptographic systems is often based on some computational assumptions such as
 the hardness of factoring of large composite numbers
 \cite{Applied_Cry,Num_Theo}. Remarkably, in 1994 Shor found an efficient
 {\em quantum} algorithm for factoring \cite{Shor94,Shor95}. Consequently, much
 of the conventional cryptography will fall apart, if a quantum computer is
 ever built.
\par
 Interestingly, it has been proposed that quantum mechanics also comes to the
 rescue. In quantum mechanics, there is a well-known ``no-cloning theorem''
 saying that an unknown quantum state cannot be cloned
 \cite{No-Cloning,No-Cloning2}. Consequently, eavesdropping in the quantum
 world will, in general, disturb the quantum state one is listening to. Thus,
 an eavesdropper can be discovered readily. Bennett and Brassard had shown in
 1984 how quantum cryptography can be used to secure communications between two
 users against eavesdropping attack through the so-called quantum key
 distribution scheme \cite{BB84}. This article does {\em not} concern quantum
 key distribution. Instead, we concentrate on a class of more fancy schemes,
 which are probably more useful in peacetime. The basic theme in those
 applications is the protection of private information during a public
 decision.
\par
 More concretely, in today's world, sometimes we need to cooperate or negotiate
 with other people without trusting them completely. An example is
 long-distance (e.g. over the phone) coin flip . Suppose a divorced couple
 wants to decide who keeps the house by a fair coin flip. Nevertheless, they no
 longer trust each other. The problem is, therefore, how this can be done
 fairly without having to arrange a meeting or to trust a third party to flip
 the coin.
\par
 Before addressing the above problem, let us consider a simpler scheme. Suppose
 Alice has chosen a number either zero or one. And she wants to give Bob a
 piece of evidence that she has made up her mind in such a way that (i) Bob
 knows nothing about Alice's choice at this moment; and (ii) Alice can no
 longer change her mind without being caught by Bob when she publicly announces
 her choice at a later time. This kind of task is called bit commitment
 \cite{Applied_Cry1}.
\par
 Clearly, bit commitment can be used to achieve coin tossing. Alice commits to
 a bit --- zero or one. Then Bob guesses which bit Alice has chosen. Finally,
 Alice opens her commitment by telling Bob which bit she has chosen. Bob
 verifies that Alice has been honest in executing the scheme. It turns out that
 bit commitment is a very important primitive in cryptography
 \cite{Applied_Cry,Kilian}. As will be discussed in later Sections, the
 security of conventional bit commitment usually relies on computational
 assumptions which can be broken in theory by exhaustive computer analysis.
 There had been a widespread belief that {\em quantum} schemes can get rid of
 computational assumptions, thus solving a long standing problem in
 cryptography.
\par
 The main focus of this review is the surprising result that this widespread
 belief has been misplaced. If Alice has a quantum computer, she can make an
 empty promise to Bob (i.e., Alice can change her choice at any time before she
 publicly opens her commitment) without being caught. This discovery
 represents a major victory of quantum cryptanalysts (i.e., code-breakers) over
 quantum cryptographers (i.e., code-makers). Finally, we remark that secure
 data transmission using quantum mechanics through the so-called quantum key
 distribution is unaffected by this new discovery.
\section{Bit Commitment --- From The Ancient To The Post-Modern World}
\label{S:BitCom}
\subsection{Bit Commitment In The Ancient World}
\label{SS:BitCom_Stone}
 The first bit commitment scheme in history probably goes as follows: First,
 Alice writes down her choice on a piece of paper, puts it in a box, and locks
 it up. She gives the box to Bob, but keeps the key herself. Later on, she
 proves her commitment to Bob (which is called opening her commitment) by
 sending the key to Bob, who can then open the box and verify the value of her
 committed bit. Although this method is simple and straight-forward, there is a
 serious loophole. The security of this simple bit commitment scheme relies
 heavily on the physical security of the box and the lock. This is clearly not
 very useful in the electronic age.
\subsection{Bit Commitment In The Modern World}
\label{SS:BitCom_Modern}
 Modern (non-quantum) bit commitment schemes rely on the idea of a one-way
 function --- a function that is easy to compute, but very hard to reverse.
 For instance, multiplying two integers is easy, but there is no known
 {\em efficient} classical algorithm\footnote{That is, an algorithm working on
 a classical computer.} to date for computing the factors of a large composite
 number \cite{Num_Theo}.
\par
 In the modern world, a bit commitment scheme may go as follows (see
 Ref.~\cite{Applied_Cry1} for discussions of various bit commitment schemes):
\par\medskip
{\bf [Classical Bit Commitment Scheme]}
\begin{enumerate}
 \item Alice chooses her bit $b=0$ to be committed to Bob. \label{I:BC_1}
 \item If $b=0$, she picks a random even number $x$ and computes $y=f(x)$ where
  $f$ is a one-way function. Similarly, if $b=1$, she picks a random odd number
  $y$ and computes $y=f(x)$. She sends $y$ to Bob. This completes the commit
  phase. \label{I:BC_2}
 \item To open her commitment, Alice sends $x$ to Bob. \label{I:BC_3}
 \item Bob verifies that $y=f(x)$ and checks whether $x$ is odd or even. This
  verifies Alice's honesty. \label{I:BC_4}
\end{enumerate}
\par\indent
 The above bit commitment scheme (as well as all other variations) relies on
 the assumption that $f^{-1}$ is hard to compute.\footnote{Actually,
 we are making a stronger assumption---that it is computationally
 infeasible to determine whether the pre-image of $f$ is even or
 odd---than the one-way function hypothesis.}
  Consequently, although Bob
 has received $y = f(x)$ in Step~\ref{I:BC_3}, he cannot invert the function
 $f$ efficiently enough to get $x$ and hence $b$ in time. In other words, even
 though Bob has all the information he needs to compute $b$ (and hence to know
 Alice's choice) before she opens her commitment, the hardness to compute
 $f^{-1}$ effectively prevents him from doing so.
\par
 Nevertheless, no one has proven the existence of a one-way function
 \cite{One-way}. Therefore, the security in this kind of bit commitment scheme
 is based on computational {\em assumptions}, which can be in principle broken
 either by exhaustive computer analysis, or by using more efficient algorithms.
\par
 To make the situation even worse, in 1994 Shor discovered an efficient quantum
 mechanical algorithm for factoring composite numbers \cite{Shor94,Shor95,RMP}.
 His algorithm makes use of the quantum interference effect and massive quantum
 parallelism in quantum mechanics, which do not have any classical counterpart.
 Since it is a technological challenge to actually build a quantum computer,
 Shor's result does not threaten classical bit commitment schemes immediately.
 However, the construction of a quantum computer is not forbidden at all by the
 laws of physics. One can envisage one day when quantum computer becomes a
 reality. Then, all classical bit commitment schemes will be unsafe.
\subsection{Bit Commitment In The Post-Modern World}
\label{SS:BitCom_Post}
 Following the pioneering works by Wiesner on ``quantum money'' and
 ``multiplexing channel'' \cite{Q_Money}, various quantum bit commitment
 schemes have been proposed \cite{BB84,BCJL,BC1,BC2}. There was a common belief
 just two years ago that quantum bit commitment is {\em absolutely} safe
 \cite{BCJL,Believe1,Believe2}. That is to say, even if both Alice and Bob have
 infinite computational power and can invoke quantum computers, any dishonest
 party will still be caught. The confidence on the security of quantum bit
 commitment is perhaps partly based on the following fact: if you are given a
 single unknown quantum state, then there is no way for you to tell exactly
 what that quantum state is. This is because measurement on an unknown quantum
 state is an irreversible process.
\par
 A number of quantum bit commitment schemes have been proposed
 \cite{BB84,BCJL,BC1,BC2}. Amongst them, the most well-known one is probably
 the BCJL scheme\cite{BCJL}. The detailed procedure of the BCJL scheme is
 irrelevant for our discussion. Nonetheless, for completeness, it is listed
 below.
\par\medskip
{\bf [BCJL Quantum Bit Commitment Scheme]}
\begin{enumerate}
 \item Let $\epsilon$ be the average noise level of a quantum communication
  channel shared between Alice and Bob. Bob chooses a Boolean matrix $G$ as the
  generating matrix of a binary linear $(n,k,d)$-code $C$ such that the ratio
  $d/n > 10\epsilon$ and the ratio $k/n = 0.52$ and announces it to Alice.
  \label{I:BCJL_1}
 \item Alice chooses a non-zero random $n$-bit string $r$ and announces it to
  Bob. \label{I:BCJL_2}
 \item Alice chooses a random $n$-bit codeword $c$ from $C$ such that the
  scalar product modulo two (i.e., the parity of the bitwise logical AND)
  between $c$ and $r$ is equal to the bit to which she is committed.
  \label{I:BCJL_3}
 \item Alice picks a random $n$-bit string $b$. Suppose the $i$th bit of $b$,
  $b_i$, equals zero. Then she sends Bob her $i$th photon in the $0^\circ$ or
  $90^\circ$ polarization according to whether $c_i = 0$ or $c_i = 1$.
  Similarly, if $b_i = 1$, she sends Bob her $i$th photon in the $45^\circ$ or
  $135^\circ$ polarization according to whether $c_i = 0$ or $c_i = 1$.
  \label{I:BCJL_4}
 \item Bob chooses a random $n$-bit string $b'$. He measures the $i$th photon
  that he receives from Alice in the $0^\circ$ and $90^\circ$ polarization
  basis if $b_i' = 0$. Otherwise, he measures the $i$th photon using the
  $45^\circ$ and $135^\circ$ polarization basis. In either case, he writes down
  the measurement results. \label{I:BCJL_5}
 \item To open her commitment, Alice reveals $c$, $b$ and her committed bit to
  Bob. \label{I:BCJL_6}
 \item Bob verifies that $c$ is a codeword. Also, if both Alice and Bob use the
  same basis for transmission and measurement, then their results $c_i$ and
  $c_i'$ must agree in the absence of noise. Therefore, Bob verifies that the
  error rate in these cases is less than the acceptable value of $1.4\epsilon$.
  Finally, Bob checks that the parity of the scalar product modulo two between
  $r$ and $c$ is indeed Alice's committed bit. Bob accepts Alice's commitment
  only if Alice passes all the three tests above. \label{I:BCJL_7}
\end{enumerate}
\par\indent
 In spite of its apparent complexity, the essential idea behind the BCJL scheme
 can be readily understood. Alice encodes her commitment as some polarization
 of photons that is unknown to Bob. Thus, it is impossible for Bob to determine
 Alice's choice before she opens her commitment. Indeed, Brassard {\em et
 al.} \cite{BCJL} have already proven the security of the BCJL scheme against a
 cheating Bob. The alleged security of this scheme against a cheating Alice is,
 however, flawed. Mayers \cite{Mayers} and, independently, we ourselves
 \cite{Lo_Chau} showed that Alice can cheat successfully if she has a quantum
 computer. As it turns out, the same cheating strategy can break not only all
 the existing schemes, but also {\em all} quantum bit commitment schemes
 \cite{Mayers1,Mayers2,Multi_Party1} that one can possibly construct. So, let
 us tell you what the most general form of quantum bit commitment scheme is
 before proving that it is necessarily insecure.
\section{Insecurity Of Quantum Bit Commitment}
\label{S:Insecure_Bit_Com}
\subsection{General Form Of A Quantum Bit Commitment Scheme}
\label{SS:Form_Bit_Comm}
 As will be argued in Subsection~\ref{SS:Generality} below, when appropriately
 formulated, the most general form of a quantum bit commitment scheme goes as
 follows \cite{Mayers,Lo_Chau,Mayers1,Mayers2,Multi_Party1,Multi_Party2}:
\par\medskip
{\bf [General Quantum Bit Commitment Scheme]}
\begin{enumerate}
 \item Alice and Bob both initialize the quantum particles at their hands to
  a prescribed state. \label{I:Gen_BC_1}
 \item Alice applies a unitary transformation to the quantum particles at her
  hand according to the value of her committed bit. Then she sends some of her
  quantum particles to Bob. \label{I:Gen_BC_2}
 \item After receiving the quantum particles from Alice, Bob applies a unitary
  transformation to the quantum particles at his hand. He then sends some of
  his quantum particles to Alice. \label{I:Gen_BC_3}
 \item Steps~\ref{I:Gen_BC_2} and~\ref{I:Gen_BC_3} are repeated finite number
  of times. \label{I:Gen_BC_4}
 \item To open her commitment, Alice sends all her particles to Bob.
  \label{I:Gen_BC_5}
 \item After receiving Alice's particles, Bob measures the composite system to
  verify Alice's honesty. \label{I:Gen_BC_6}
\end{enumerate}
\subsection{Unitary Description}
\label{SS:Math_Form}
 Let us formulate the above description in mathematics.\footnote{Mayers proved
 that all quantum bit commitment schemes are insecure in
 Refs.~\cite{Mayers1,Mayers2}. Here we will, however, follow our discussion of
 the same result in Refs.~\cite{Multi_Party1}.} Our justifications that the
 scheme is general will be made in Subsection~\ref{SS:Generality}. Let us
 denote the Hilbert spaces of Alice's and Bob's quantum machines by $H_A$ and
 $H_B$, respectively. And the Hilbert space of the quantum communication
 channel is denoted by $H_C$. A quantum bit commitment scheme is executed in
 $H = H_A \otimes H_B \otimes H_C$. Initially, Alice prepares a state $|0
 \rangle_A$ or $|1\rangle_A$ according to the value that she would like to be
 committed to Bob. Bob prepares a fixed state $|v\rangle_B$ for $H_B \otimes
 H_C$. This is Step~\ref{I:Gen_BC_1} of the general scheme. Consequently, the
 initial state is $|u_b \rangle = |b\rangle_A \otimes |v\rangle_B$ when Alice
 is committed to $b$ ($b = 0,1$). The two parties now take turns to perform
 unitary transformations (Steps~\ref{I:Gen_BC_2}--\ref{I:Gen_BC_4}). That is,
 in each step, a party $D\in\{A, B\}$ applies a unitary transformation on the
 system $H_D \otimes H_C$. Such a unitary transformation induces a unitary
 transformation on the larger space $H$.
\par
 The upshot is that the whole procedure of the commit phase, being a sequence
 of unitary transformations on $H$, can be summarized by a single unitary
 transformation $U$ applied to the initial state on $H$. Such a unitary
 description will greatly simplify our discussion: At the end of the commit
 phase, Alice and Bob share a pure state, either $U\left( |0\rangle_A \otimes
 |v\rangle_B \right)$ or $U \left( |1\rangle_A \otimes |v\rangle_B \right)$.
 Also, since Alice and Bob know the procedure of the protocol, they also know
 $U$. So, once Alice opens her commitment by sending all her particles to Bob
 (Step~\ref{I:Gen_BC_5}), Bob can readily verify Alice's claim
 (Step~\ref{I:Gen_BC_6}).
\par
 Here we assume the most advantageous situation for Bob where during the
 opening phase Alice sends {\em all} her particles to Bob. We shall show that
 even then Alice can cheat successfully.
\subsection{Generality Of The Above Description}
\label{SS:Generality}
 Let us explain why the BCJL protocol falls into the above general scheme.
 Clearly, except for the selection of the error correcting code in
 Step~\ref{I:BCJL_1}, the BCJL protocol involves only one way communications
 from Alice to Bob. Also, sending photons with different polarization to Bob in
 Step~\ref{I:BCJL_4} of the BCJL scheme is equivalent to first applying a
 unitary transformation to the initialized photons by Alice before sending them
 to Bob. Moreover, it does no harm for Bob to delay his measurement in
 Step~\ref{I:BCJL_5} of BCJL until Alice opens her commitment.
\par\indent
 At this point, readers may question if the above commitment scheme is the most
 general one. In particular, they may raise the following objections:
\newcounter{qa}
\setcounter{qa}{1}
\begin{list}{}{\setlength{\itemindent}{0in}
 \settowidth{\labelwidth}{\rm Question 3:} \setlength{\leftmargin}{\labelwidth}
 \addtolength{\leftmargin}{\labelsep}}
 \item[Question \arabic{qa}:] Communications by classical means between Alice
  and Bob is not considered.
 \item[Answer \arabic{qa}:] Since classical communications are a special case
  of quantum communications, they can be done on a quantum channel and there is
  no need to give them any special consideration. 
 \newcounter{QA:1} \setcounter{QA:1}{\value{qa}}
\addtocounter{qa}{1}
 \item[Question \arabic{qa}:] Alice and Bob may measure some of their quantum
  particles in Steps~\ref{I:Gen_BC_2}--~\ref{I:Gen_BC_4}. Moreover, the unitary
  transformations they apply may depend on the results of their measurements.
 More importantly, measurements
 give rise to {\it decoherence}.
 Wouldn't a bit commitment scheme with some measurements be secure?
 \newcounter{QA:2} \setcounter{QA:2}{\value{qa}}
 \item[Answer \arabic{qa}:] Alice and Bob can delay their measurements until
  the opening of the commitment. For example, given a bit commitment scheme
  that involves a measurement by Alice and that Alice is supposed to apply a
  unitary transformation $U_i$ to the rest of her quantum particles if her
  measurement result is $|e_i\rangle$ for some $i$. She can define another
  linear operator $U$ which maps $|e_i\rangle \otimes |\Psi\rangle$ to $|e_i
  \rangle \otimes U_i |\Psi\rangle$ for each $i$. Clearly, $U$ is a unitary
  operator. Therefore, Alice may choose to apply $U$ to her quantum particles
  and {\em delay} her measurement until the opening phase.
 
  Even bit commitment schemes with measurements are insecure.
  The key insight is the following: To show that
 {\it all} bit commitment schemes (classical, quantum or quantum
 but with some measurements) are insecure, it suffices to
 consider {\it only} a general fully quantum bit commitment scheme where both
  Alice and Bob have quantum computers. This is because {\it any} other
  procedure followed by Bob in a bit commitment scheme can be
 rephrased as a quantum bit commitment scheme where Bob does
 have a quantum computer but just fails to make full use of it.

  Now, we will show that Alice has a winning strategy against Bob
 even if he makes full use of his quantum computer. It is then clear
 that this ``sure-win'' strategy by Alice will defeat a Bob who fails to
 make full use of his quantum computer. Therefore, the insecurity of a
 fully quantum bit commitment scheme automatically implies the
 insecurity of all bit commitment schemes (purely quantum, classical or
 quantum scheme but with measurements).

 Notice also that a cheating Alice generally needs a quantum computer to cheat.
\addtocounter{qa}{1}
 \item[Question \arabic{qa}:] Alice and Bob may throw dice to decide which
  unitary transformation to use. Moreover, they may invoke ancillary quantum
  particles. More generally, Alice and Bob are dealing with density matrices,
  not wavefunctions.
 \newcounter{QA:3} \setcounter{QA:3}{\value{qa}}
 \item[Answer \arabic{qa}:] Using the same argument in Answer~\arabic{QA:2},
  Alice and Bob can delay the throwing of the die (i.e., the state of the die
  is kept in a quantum superposition and does not collapse) until the opening
  phase. Any ancilla (including the quantum die) can be incorporated into
  Alice and Bob's quantum machines right at the beginning. This simply leads to
  an extension of the dimensions of the Hilbert spaces $H_A \otimes H_B$.
  Moreover, the state in the tensor product of these extended Hilbert spaces
  is {\em pure}.
\addtocounter{qa}{1}
 \item[Question \arabic{qa}:] Instead of manipulating the quantum particles in
  turn, Alice and Bob may manipulate, send out, and receive their quantum
  particles in parallel. That is to say, Steps~\ref{I:Gen_BC_2}
  and~\ref{I:Gen_BC_3} are space-like events.
 \newcounter{QA:4} \setcounter{QA:4}{\value{qa}}
 \item[Answer \arabic{qa}:] In practice, it is impossible to ensure that Alice
  and Bob receive signals simultaneously. This is because, for two distant
  observers, there is no way for one to be sure of the physical location of the
  other. (Recall that one of the two persons may be cheating.) More
  importantly, simultaneity has no invariant meaning in special relativity.
\end{list}
\par\noindent
 Having convinced ourselves that the above bit commitment scheme is the most
 general one, we turn to Alice's cheating strategy. First, we need a technical
 result.
\subsection{Schmidt Decomposition\protect\footnote{The discussion in this Subsection is
 based on Ref.~\protect\cite{Schmidt}.}~}
\label{SS:Schmidt_Decomp}
 Let $H_A$ and $H_B$ be Hilbert spaces with dimensions $p$ and $q$,
 respectively. And let $|\Phi\rangle$ be any normalized state in $H_A \otimes
 H_B$. Define the density matrix $\rho = |\Phi\rangle\langle\Phi |$, and
 reduced density matrices $\rho^A = \mbox{Tr}_B \rho$ and $\rho^B =
 \mbox{Tr}_A \rho$.
\par\medskip\noindent
{\it Claim}: $|\Phi\rangle$ can be written as
\begin{equation}
 |\Phi\rangle = \sum_{i=1}^r \sqrt{\lambda_i} | a_i \rangle \otimes | b_i
 \rangle ~, \label{E:Schmidt_Polar}
\end{equation}
 where $| a_i \rangle$ and $| b_i \rangle$ are orthonormal eigenstates of
 $\rho^A$ and $\rho^B$, respectively. In addition, $r\leq\min (p,q)$ is the
 total dimension of the non-zero eigenspaces of $\rho^A$. This representation
 is called the Schmidt decomposition \cite{Schmidt}.
\par\medskip\noindent
 {\it Proof:} Any $|\Phi\rangle$ can be written in terms of the orthonormal
 eigenbasis $\{ | a_i \rangle \}$ of $\rho^A$ as
\begin{equation}
 |\Phi\rangle = \sum_{i=1}^p | a_i \rangle \otimes | b'_i \rangle ~,
 \label{E:Phi_1}
\end{equation}
 where $| b'_i \rangle$'s are not necessarily
 orthogonal. By taking a trace over $H_B$, we find
\begin{eqnarray}
 \mbox{Tr}_B |\Phi\rangle\langle\Phi | & = & \mbox{Tr}_B \sum_{i=1}^p
 \sum_{j=1}^p | a_j \rangle \otimes | b'_j \rangle \langle a_i | \otimes
 \langle b'_i | \nonumber \\
& =&    \sum_{i=1}^p
 \sum_{j=1}^p \sum_{k =1}^q \langle \hat{b}_k | b'_j \rangle 
| a_j \rangle \langle a_i |
 \langle b'_i |  \hat{b}_k \rangle \nonumber \\
& =& \sum_{i=1}^p
 \sum_{j=1}^p \sum_{k =1}^q \langle b'_i |  \hat{b}_k \rangle
\langle \hat{b}_k | b'_j \rangle | a_j \rangle \langle a_i | \nonumber \\
& = & \sum_{i=1}^p \sum_{j=1}^p \langle b'_i |
  b'_j \rangle | a_j \rangle \langle a_i | ~, \label{E:Trace_B}
\end{eqnarray}
where $ |  \hat{b}_k \rangle$'s form an orthonormal basis in $H_B$.
 Equating this to
\begin{equation}
 \rho^A = \sum_{i=1}^r \lambda_i | a_i \rangle \langle a_i |
 \label{E:rho_A_Sch}
\end{equation}
 gives $\langle b'_i | b'_j \rangle = \lambda_i \delta_{ij}$. Hence, $| b_i
 \rangle = \lambda_i^{ - { 1 \over 2}}
 | b'_i \rangle $ is an orthonormal set in $H_B$,
 and the Schmidt decomposition in Eq.~(\ref{E:Schmidt_Polar}) holds.
\par
 Similarly, by taking the trace of $|\Phi \rangle$ over $H_A$, we arrive at
\begin{equation}
 \rho^B = \sum_{i=1}^r \lambda_i | b_i \rangle \langle b_i | ~.
 \label{E:rho_B_Sch}
\end{equation}
 Therefore, $| b_i \rangle$ is an eigenvector of $\rho^B$ corresponding to the
 eigenvalue $\lambda_i$.
\par\noindent
{\it Q.E.D.}
\subsection{Alice's cheating strategy}
\label{SS:Alice_BC_Cheat}
 Now, we show that the two basic security requirements of quantum bit
 commitment are inconsistent: In fact, if Bob cannot learn the value of the
 committed bit $b$, then Alice can almost always cheat successfully by changing
 the value of $b$ at the beginning of the opening phase without being caught by
 Bob.
\par
 Consider the combined quantum state of the particles in Alice and Bob's hand
 just before the opening phase. We can include $H_C$ to the quantum machine of
 whoever controlling the channel at this point. Therefore, $H=H_A \otimes H_B$
 simply. When the committed bit, $b$, is zero, the state of the composite
 system can be written in Schmidt decomposition (see
 Eq.~(\ref{E:Schmidt_Polar})) as
\begin{equation}
 |0_{\rm final}\rangle = \sum_i \sqrt{\alpha_i} |e_i\rangle_A \otimes |\phi_i
 \rangle_B ~. \label{E:Commit_0}
\end{equation}
 On the other hand, when the committed bit, $b$, is one, it can be written in
 Schmidt decomposition as
\begin{equation}
 |1_{\rm final}\rangle = \sum_i \sqrt{\beta_i} |e_i'\rangle_A \otimes |\phi'_i
 \rangle_B ~. \label{E:Commit_1}
\end{equation}
\par\indent
 The quantum state of Bob's particles, without the extra information coming
 from Alice, can be described by a density matrix obtained by taking a partial
 trace of the entire wavefunction over the particles at Alice's hand. If $b=0$,
 Bob's density matrix is
\begin{equation}
 \mbox{Tr}_A ( |0_{\rm final}\rangle\langle 0_{\rm final}| ) \equiv \rho^B_0 =
 \sum_i \alpha_i |\phi_i\rangle_B \langle \phi_i |_B ~. \label{E:Bob_0}
\end{equation}
 Similarly, if $b =1$, Bob's density matrix is
\begin{equation}
 \mbox{Tr}_A ( |1_{\rm final}\rangle\langle 1_{\rm final}| ) \equiv \rho^B_1 =
 \sum_i \beta_i |\phi'_i\rangle_B \langle \phi'_i |_B ~. \label{E:Bob_1}
\end{equation}
\par\indent
 In order that Bob has little chance to know Alice's choice in advance, we
 require the reduced matrices $\mbox{Tr}_A ( |0_{\rm final}\rangle \langle
 0_{\rm final}| ) \equiv \rho^B_0$ and $\mbox{Tr}_A ( |1_{\rm final}\rangle
 \langle 1_{\rm final}| ) \equiv \rho^B_1$ to be as ``close'' as possible.
\par
 Let us first consider the {\em ideal} case when $\rho^B_0 = \rho^B_1$. It then
 follows\footnote{Here we assume that the eigenstates are non-degenerate. The
 case of degenerate eigenstates can be dealt with in a similar manner.} from
 Eqs.~(\ref{E:Bob_0}) and~(\ref{E:Bob_1}) that
\begin{equation}
 \alpha_i = \beta_i \label{E:coeff}
\end{equation}
 and
\begin{equation}
 |\phi_i\rangle_B = |\phi'_i\rangle_B ~. \label{E:state}
\end{equation}
 for all $i$. Substituting Eqs.~(\ref{E:coeff}) and~(\ref{E:state}) into
 Eq.~(\ref{E:Commit_1}), we get
\begin{equation}
 |1_{\rm final}\rangle = \sum_i \sqrt{\alpha_i} |e_i'\rangle_A \otimes |\phi_i
 \rangle_B ~. \label{E:Commit_1'}
\end{equation}
\par\indent
 Let us consider the unitary transformation $U^A $ which maps $|e_i\rangle_A$
 to $|e_i'\rangle_A$. Notice that it is a {\em local} unitary transformation by
 Alice and as such can be implemented by Alice {\em alone}. Remarkably, it maps
 $|0_{\rm final}\rangle $ to $|1_{\rm final}\rangle$. In other words, Alice can
 always cheat by changing her bit from $0$ to $1$ just before she opens her
 commitment. More concretely, the cheating strategy goes as follows: She always
 executes the protocol for $b=0$ during the commitment phase. At the beginning
 of the opening phase, she decides on the value of $b$ that she would like to
 open. Suppose she decides $b=0$ now, she simply executes the protocol
 honestly. On the other hand, if she now chooses $b=1$, she applies $U^A$ to
 her state. This changes $|0_{\rm final}\rangle $ to $|1_{\rm final}\rangle$.
 She can then declare that $b=1$ and execute the opening phase for $b=1$
 accordingly. There is absolutely no way for Bob to defeat such an attack by
 Alice.
\par
 Having considered the ideal case, let us now, following Mayers \cite{Mayers},
 consider the non-ideal case where $\rho^B_0$ differs from $\rho^B_1$ slightly.
 In quantum mechanics, a good measure of the ``closeness'' between two density
 matrices is fidelity \cite{Fidelity}. In general, given two reduced density
 matrices $\rho_0^B$ and $\rho_1^B$ of Bob, there are many possible systems A
 attached to Bob's system B such that the {\em combined} wavefunction of
 systems A and B are pure states $|\Psi_0\rangle$ and $|\Psi_1\rangle$,
 respectively. That is, $\mbox{Tr}_A ( |\Psi_i\rangle\langle \Psi_i| ) =
 \rho_i^B$ for $i = 0, 1$. This kind of pure states $|\Psi_i\rangle$ are called
 {\em purifications}. The fidelity can be defined as
\begin{equation}
 F(\rho_0^B, \rho_1^B) = \max \left( \left| \langle\Psi_0 |\Psi_1 \rangle
 \right| \right) ~, \label{E:Fidelity_Def}
\end{equation} 
 where the maximization is taken over all possible purifications. Clearly $0
 \leq F \leq 1$. Moreover, $F = 1$ if and only if there is a purification
 such that $|\Psi_0\rangle = |\Psi_1\rangle$, which in turn holds if and only
 if $\rho_0^B = \rho_1^B$. The closer the two reduced density matrices, the
 higher their fidelity.
\par
 Therefore, the requirement that Bob has little chance to know Alice's choice
 in advance implies that
\begin{equation}
  F (\rho_0^B, \rho_1^B )=  1 - \delta \label{E:Fidelity_Bob}
\end{equation}
 for some small $\delta \geq 0$.
\par
 Here come two simple but crucial remarks. First, for any fixed purification
 $|\Psi_1\rangle$ of $\rho_1^B$, there exists a maximally parallel purification
 $|\Psi_0\rangle$ of $\rho_0^B$ such that Eq.~(\ref{E:Fidelity_Def}) is
 satisfied. Second, it can be proved that any two purifications $|\Psi_0
 \rangle$ and $|\Psi'_0 \rangle$ of the same density matrix $\rho_0^B$ are
 necessarily related by a {\em local} unitary transformation by Alice
 {\em alone}. These two facts follow trivially from the form of Schmidt
 decomposition in Eq.~(\ref{E:Schmidt_Polar}).
\par
 Let us apply these two remarks to non-ideal quantum bit commitment. From the
 first remark, given a purification $|1_{\rm final}\rangle$ of $\rho_1^B$ in
 Eq.~(\ref{E:Commit_1}), there exists a purification $|0' \rangle$ of
 $\rho_0^B$ such that
\begin{equation}
 \langle 0' |1_{\rm final}\rangle = 1 - \delta ~. \label{E:maximal}
\end{equation}
 From the second remark, there exists a {\em local} unitary transformation say
 $U^A$ that maps $|0_{\rm final}\rangle $ to $|0' \rangle$.
\par
 Now it is clear that, using the same cheating strategy as in the ideal case,
 Alice can almost always cheat successfully. In more detail, Alice's cheating
 strategy goes as follows: Alice chooses $b=0$ and executes the commit phase
 honestly. During the opening phase, Alice decides the value of $b$ to be
 opened. If she chooses it to be $0$, she acts honestly. However, if she
 chooses it to be $1$, she claims that $b=1$ and applies the local unitary
 transformation $U^A$ to change $|0_{\rm final}\rangle $ to $|0' \rangle$. From
 Eq.~(\ref{E:maximal}), it is very hard for Bob to distinguish the state in the
 dishonest case, $|0' \rangle$, from the state in the honest case,
 $|1_{\rm final}\rangle$. Therefore, Alice can almost always cheat
 successfully.
\par
 Notice that the cheating strategy makes essential use of entanglement. To
 succeed in cheating, Alice must be able to store quantum signals for a long
 time and to coherently manipulate quantum particles. That is, Alice generally
 needs a quantum computer.
\par
 At this moment, readers may ask why the no-cloning theorem and uncertainty
 principle cannot prevent Alice from cheating. The reason is simple: It is
 impossible for Bob to verify every unitary transformation and measurement
 that Alice has made. Therefore, Alice can delay making her unitary
 transformation $|0_{\rm final}\rangle \longrightarrow |0'\rangle$ till the
 opening phase.
\section{Concluding Remarks}
\label{S:Conclude}
\subsection{Secure Computations}
\label{SS:Secure_Comp}
 Quantum bit commitment is a basic building block for many other quantum
 cryptographic protocols. After the fall of quantum bit commitment, the
 security of other quantum two-party protocols, in particular, the so-called
 two-party secure computations also came into question.
\par
 In a {\em one-sided} two-party secure computation, Alice with a secret $x$ and
 Bob with a secret $y$ would like to cooperate to compute a prescribed function
 $f(x,y)$ such that at the end, (i) Alice learns nothing (about $y$ and
 $f(x,y)$); (ii) Bob learns $f(x,y)$; and (iii) Bob learns nothing about $x$
 except for what logically follows from $y$ and $f(x,y)$.
\par
 One-sided two-party secure computations can, for instance, be used to prevent
 a fake teller machine from stealing a customer's PIN (Personal
 Identification Number): To do this, let $x$ be the customer's (i.e., Alice's)
 PIN , $y$ be the record of the customer's PIN in the teller machine (i.e.,
 Bob). Consider the function $f(x,y) = \delta_{xy}$. Running the one-sided
 two-party computation of $f(x,y)$ will allow the teller machine to verify
 whether the customer's input $x$ matches the record $y$ of the teller machine.
 However, a fake teller machine does not know which $y$ to use as the input.
 Using a random $y$ will give it very little information about $x$.
\par
 The insecurity of quantum one-sided two-party secure computations was finally
 demonstrated explicitly by one of us \cite{Multi_Party2}, who showed that a
 cheating Bob can learn $f(x,y)$ for {\em all} values of $y$. This is a fatal
 violation of the security requirement. For instance, in the above password
 verification scheme such a cheating Bob will, by testing all possible values
 of $y$, learn the customer's input $x$.
\par
 A cheating Bob proceeds as follows: Bob inputs $y=y_1$, executes the protocol
 honestly and learns $f(x,y_1)$ by performing a measurement. He then applies a
 unitary transformation to change the value of $y$ from $y_1$ to $y_2$ and
 learns $f(x,y_2)$ by performing a measurement. After that, he applies a
 unitary transformation to change the value of $y$ from $y_2$ to $y_3$ to learn
 $f(x,y_2)$ and so on.
\par
 This cheating strategy works chiefly for two reasons. First, the measurement
 of say $f(x,y_1)$ in no way disturbs the state under observation. This is so
 because the state is an {\em eigenstate} of $f(x,y_1)$.\footnote{This is
 because Bob is supposed to be able to determine $f(x,y_1)$ unambiguously. Here
 we are considering the ideal case. The non-ideal case where the state is only
 approximately an eigenstate of $f(x,y_1)$ does not change the essential
 argument \cite{Multi_Party2}.} Second, the essence of the insecurity of
 quantum bit commitment is that if a party $A$ knows nothing about the input
 $b$ of another party $B$ even at the end of the protocol, then $B$ can cheat
 by changing $b$ at the very end. Now since in a one-sided two-party secure
 computation Alice cannot learn about $y$, a cheating Bob can change the value
 of $y$. That is, the state of all quantum particles in Alice and Bob's hands
 when computing $f(x,y)$ and $f(x,y')$ are related by a unitary transformation
 involving only particles in Bob's hand \cite{Multi_Party2}, as required in the
 cheating strategy presented in the last paragraph.
\par
 In conclusion, quantum one-sided two-party secure computations are, {\em in
 principle}, insecure.\footnote{Another interesting protocol is quantum coin
 tossing, we have shown in Ref.~\cite{Multi_Party1} that {\em ideal} quantum
 coin tossing (that completely forbids successful cheating) is impossible. It
 is still open whether non-ideal coin tossing is achievable. It was also shown
 in Ref.~\cite{Multi_Party2} that quantum {\em two-sided} two-party secure
 computations are also generally impossible.} Even though quantum bit
 commitment and quantum two-party secure computations are insecure {\em in
 theory}, they may still be secure {\em in practice}. This is because a cheater
 generally needs a quantum computer to cheat successfully. And it is a
 technological feat to build a quantum computer. The implication is that, by
 working with quantum protocols, one may replace classical computational
 assumptions with quantum computational assumptions.
\subsection{Security Analysis of Composite Quantum Protocols}
\label{SS:Security_Analysis}
 In the security analysis of quantum protocols, researchers usually only
 consider the case when a protocol is executed only once and in isolation. This
 is, however, contrary to the spirit that a cryptographic protocol satisfies
 conventional security requirements, which are usually written in terms of
 probability and thus implicitly demand that protocols follow the rules of
 inference in classical probability theory. Therefore, in analyzing quantum
 protocols a more refined security analysis than what is commonly adopted is
 needed \cite{Multi_Party2}. In order to be able to apply classical probability
 theory to the study of a composite protocol, it is crucial to study the
 security of quantum protocols not only when they are used in isolation, but
 also when they are used as ``black-box'' primitives in building up more
 complicated protocols. It is only when they pass such a stringent test that
 they should be certified as secure.
\par
 Of course, such security analysis may be difficult to perform in practice.
 However, this is the price that one has to pay in asserting that a quantum
 scheme achieves a set of security requirements which are written in terms of
 classical {\em probability}.\footnote{The only alternative that we can think
 of is to describe the security requirements of quantum cryptographic protocols
 in terms of {\it probability amplitude}. Such an alternative has not been
 given serious consideration so far.}
\par
 With this more stringent and, in our opinion, more accurate security
 analysis, classical inference is, by definition, valid. Since it is a standard
 result in classical cryptography that some two-party secure computations can
 be used to implement bit commitment \cite{Kilian}, the impossibility of
 quantum bit commitment must immediately imply that quantum two-party secure
 computation is generally impossible.
\subsection{Lessons We Learn}
\label{SS:Lesson}
 We remark that the attacks used by Mayers in
 Refs.~\cite{Mayers,Mayers1,Mayers2}, by Lo and Chau in
 Refs.~\cite{Lo_Chau,Multi_Party1} and by Lo in Ref.~\cite{Multi_Party2} as
 discussed in this paper, were not new. A weakness of a restricted class of
 quantum secure computation schemes (``multiplexing channel'' \cite{Q_Money})
 as well as the Einstein-Podolsky-Rosen-type of attack \cite{BB84} which
 underlines the insecurity of quantum bit commitment and secure computations
 had already been noted in some pioneering papers. What had not been fully
 appreciated until the work of Mayers \cite{Mayers,Mayers1,Mayers2} and ours
 \cite{Lo_Chau,Multi_Party1,Multi_Party2} was the {\em generality} of such
 attacks.
\par
 Quantum mechanics is a double-edged sword in cryptology. While it apparently
 equips cryptographers with secure schemes of quantum key
 distribution\footnote{Despite many interesting approaches proposed in the
 literature \cite{OT2,QKD2,QKD3,QKD4,QKD5,QKD6,QKD7}, in our opinion, a widely
 accepted complete proof of the security of quantum cryptography in a noisy
 channel is still missing.} due to the quantum no-cloning theorem, it also
 gives the quantum cryptanalyst the Einstein-Podolsky-Rosen effect which allows
 him to delay his measurement and defeat quantum bit commitment and secure
 computations. Now on one hand, we generally believe that quantum key
 distribution is secure. On the other hand, quantum bit commitment and
 one-sided two-party secure computations have been shown to be impossible. A
 natural question to ask is: What is the exact boundary to the power of quantum
 cryptography? For instance, does quantum cryptography help multi-party secure
 computations? The answers to these questions may give us new insights on
 quantum information theory.
\par
 We must emphasize that the security of quantum key distribution is unaffected
 by the attacks described in this paper. Quantum key distribution alone should
 guarantee that quantum cryptography remains a fertile subject for future
 investigations. This is so particularly because of the dramatic recent
 progress in experimental quantum cryptography
 \cite{Expt1,Expt2,Expt3,Expt4,Expt5}.
\acknowledgments
 We thank many helpful discussions with numerous colleagues including M.
 Ardehali, C.~H. Bennett, G. Brassard, C. Cr\'{e}peau, D.~P.
 DiVincenzo, L. Goldenberg, J. Hrub\'{y}, R. Jozsa, J. Kilian, D. Mayers, J.
 Preskill, P. Shor, T. Spiller, T. Toffoli, L. Vaidman and F. Wilczek after the
 completion of an earlier version of Ref.~\cite{Lo_Chau}. We also thank Sam
 Braunstein for his kind invitation to write this survey paper, Kenny
 Paterson for providing references and R. Cleve
 for pointing out some inaccuracy in an earlier version of
 this paper. One of us (H. F. C.) is supported by the
 RGC grant HKU~7095/97P.

\end{document}